\documentclass[final,3p,times,twocolumn]{elsarticle}
\usepackage{amssymb}
\journal{Physica E}

\begin{document}

\begin{frontmatter}

\title{High cumulants in the counting statistics measured for a quantum dot}

\author[luh]{Christian Fricke}
\author[luh]{Frank Hohls}
\author[harvard]{Christian Flindt}
\author[luh]{Rolf J. Haug}
\address[luh]{Institut f\"ur Festk\"orperphysik, Leibniz Universit\"at Hannover, 30167 Hannover, Germany.}
\address[harvard]{Department of Physics, Harvard University, 17 Oxford Street, Cambridge, MA 02138, USA.}

\begin{abstract}
We report on measurements of single electron tunneling through a quantum dot using a quantum point contact as non-invasive charge detector with fast time response. We elaborate on the unambiguous identification of individual tunneling events and determine the distribution of transferred charges, the so-called full counting statistics. We discuss our data analysis, including the error estimates of the measurement, and show that the quality of our experimental results is sufficiently high to extract cumulants of the distribution up to the 20th order for short times.
\end{abstract}


\end{frontmatter}

\begin{figure}[ht]
\begin{center}
\includegraphics[width=0.95\linewidth]{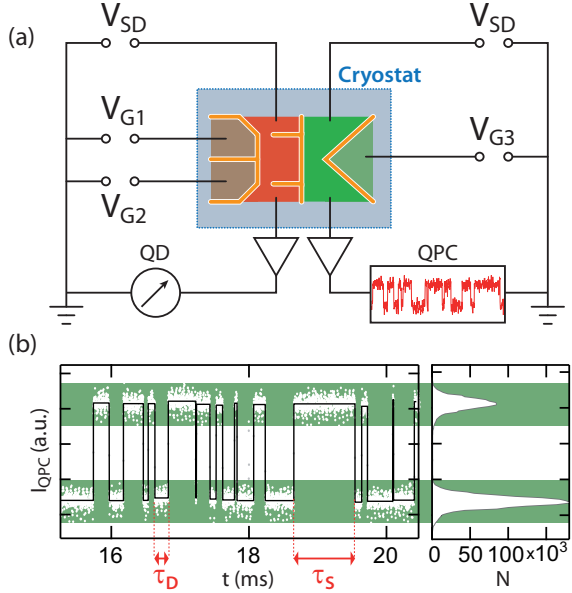}
\caption{Operating principle for measuring single electrons passing through a QD using a
QPC as detector. (a) Schematic picture of the experimental setup. Current through QPC and QD are measured in a two circuit setup. The QPC current is measured with a 100 kHz bandwidth. (b) Dots: Time
segment of the raw measured QPC current signal. Line: Resulting signal after analysis of the raw data. Tunneling times $\tau_\mathrm{S}$ and
$\tau_\mathrm{D}$ deduced from this analysis are marked by dotted lines. Histogram: Distribution of currents from the time trace. From the two peaks in the distribution, two bands are extracted corresponding to 0 (upper band) and 1 (lower band) additional electrons on the QD.} \label{uebersicht}
\end{center}
\end{figure}

The progress in time-resolved charge sensing using quantum point contacts has made the detection of individual electron tunneling events possible \cite{Elzerman2003, schleser2005, fujisawa2004}. Such experimental techniques enable measurements of the distribution of transferred charges in a quantum dot system, the so-called full counting statistics \cite{Levitov96,Gustavsson2006,Fricke2007}. Measurements of the full counting statistics have been used to study the first few higher order cumulants of the distribution  \cite{Gustavsson2007}. The first cumulant is the mean of the number of transferred electrons $n$, the second is the variance, and the third is the skewness.  Recently, the general behavior of cumulants of very high orders was investigated experimentally as well as theoretically~\cite{PNAS}.

Here, we report on real time single electron counting with a large bandwidth detector. This enables us to measure the full counting statistics for single electron transport through a quantum dot with high precision. From the distribution of tunneling events the cumulants describing the statistical properties of the system are extracted. The cumulants of the transport statistics are widely studied in theory, and here we discuss the experimental precision at which they can be measured.

Our device is based on a GaAs/AlGaAs heterostructure containing a two-dimensional electron system (2DES) 34 nm below the surface. The electron density is $\rho = 4.59 \cdot 10^{15} \hspace{1.2mm}\mathrm{m}^{-2}$, and the mobility is $\mu = 64.3 \hspace{1.2mm} \mathrm{m^2/V s}$. We have used an atomic force microscope (AFM) to define the quantum dot (QD) and the quantum point contact (QPC) structure by local anodic oxidation (LAO) on
the surface \cite{held98,keyser2000}; the 2DES below the oxidized surface is depleted and insulating areas can be written.

\begin{figure}[ht]
\begin{center}
\includegraphics[width=0.95\linewidth]{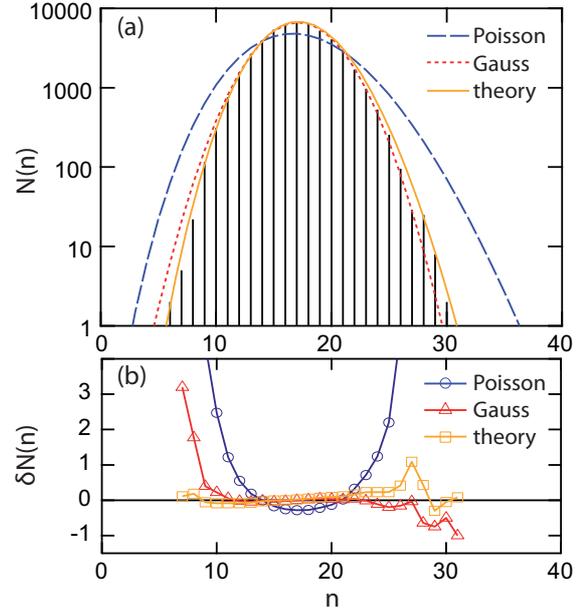}
\caption{(a) Distribution compiled from 853,181 tunneling events. The experimental result (bars) are compared with a Poisson distribution (dashed line), a Gauss distribution (dotted line), and theoretical calculations corresponding to the experimental setup (solid line). (b) Relative deviation $\delta N$ between the experimentally determined distribution and the theoretical distributions in panel (a).} \label{dist}
\end{center}
\end{figure}

A schematic view of our device is presented in Fig.~\ref{uebersicht}~(a). The bright lines depict the insulating oxide barriers written by the AFM. The QPC (right area) is separated from the QD structure (left area) by an insulating line. The QPC can be electrically tuned using an in-plane gate by applying the potential $V_{G3}$. The current through the QPC is amplified with a 100 kHz current amplifier and detected in a time-resolved manner. A small bias is chosen to avoid back-action on the QD \cite{khrapai2006}.  The QD is coupled to source and drain electrodes via two tunneling barriers, which can be separately
controlled with gate voltages $V_{G1}$ and $V_{G2}$. These gates are also used to tune the number of electrons on the QD. A sufficiently large bias $V_{SD}$ ensures that transport is unidirectional. The sample is placed in a dilution refrigerator with temperatures in the mK-regime.

The quantum dot is operated in the Coulomb blockade regime close to a charge degeneracy point, where a single electron at the time can enter the QD from the source electrode and leave via the drain. Consequently, the electron number on the QD changes back and forth between 0 and 1 additional electrons as electrons tunnel through the system. This leads to a corresponding change of the electrostatic potential at the position of the QPC. The QPC is tuned to a working point on the edge of the first conductance step, where the current through the QPC is very sensitive to the presence of localized electrons on the QD. With 1 additional electron on the QD, the current through the QPC is suppressed, and the suppression is only lifted as the electron leaves the QD via the drain electrode.

\begin{figure*}[t]
\begin{center}
\includegraphics[width=0.85\linewidth]{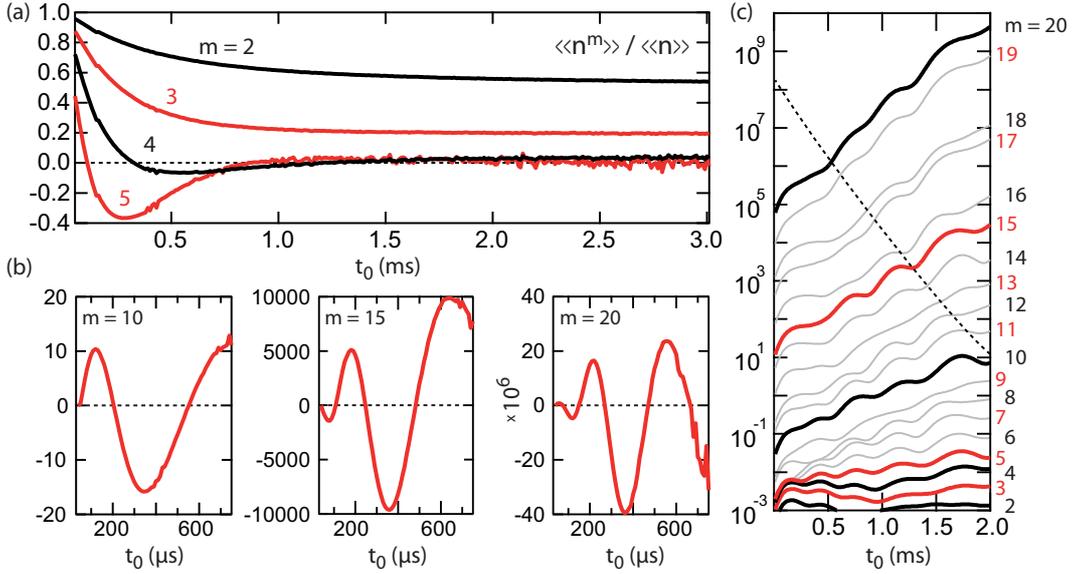}
\caption{High order cumulants (a) Normalized  cumulants $\langle\!\langle n^m\rangle\!\rangle/\langle\!\langle n\rangle\!\rangle$ for $m=2-5$. Cumulants are shown for times $t_0$ up to 3 ms. (b) Cumulants for $m=10, 15, 20$ for times $t_0$ up to 750 $\mu$s. (c) Error estimates calculated as the difference between cumulants obtained from the first and second half of the full time trace. The smoothened absolute error is shown for direct comparison with the measured cumulants. The error estimates for the measured cumulants (a, b) are shown with thick lines, while the thin lines correspond to the error estimates for other cumulants up to order 20. Where the solid lines cross the dashed line the corresponding estimated error is $\sim10\%$.} \label{cumulants}
\end{center}
\end{figure*}

The time-resolved QPC current consequently shows a random telegraph signal, as shown in Fig.\ \ref{uebersicht}~(b). The individual data points (dots) correspond to an integration time of 10 $\mu$s. Most of the data is distributed in two bands, or current levels. The histogram of the time trace on the right shows the two levels of the QPC current. The lower (upper) state corresponds to the QD with 1 (0) additional electrons. The time intervals that the QPC current stays in either of the two bands directly yield the waiting times for an electron to tunnel on ($\tau_S$) or off ($\tau_D$) the QD. The tunneling events themselves are marked by the transitions between the two bands.

In order to access the transport statistics it is necessary to extract the number of tunneling events from the raw data. This is done using an averaging algorithm. The key principle is to identify the two bands corresponding to the different charge states of the QD. To this end we use the histogram of the QPC current. The two peaks are identified and the boundaries of the two bands are located. Once the boundaries of the upper and lower band are known, tunneling events can be identified as transitions between the two current levels. We identify a transition by checking each measured current point for being part of one of the two bands. As long as the system stays in one state, the current level is averaged. In order to avoid a shifting average due to external noise sources, we neglect points that are not part of one of the two bands. If a point is located in another band than the previous point, the charge on the quantum dot has changed, and a tunneling event is registered.

The band selective averaging approach features a number of advantages. Most importantly, it is more robust to external disturbances of the measured current than the alternative method of edge detection directly at the flank of a transition. As only points inside the two bands are taken into account, noise spikes are not affecting the detection of a transition. The current averaging itself gives an additional verification, as a shift of the average between two tunneling events can be detected. In this way, unintended changes of the detector are directly observable. Also data points outside the bands can be used to identify the power of external noise added to the detector current. This allows us to identify time segments with too high noise level and it also improves the effective bandwidth of the detector.

We now turn to the statistics of the number of charges transferred through the QD. The exact times of tunneling events are stochastic, and the statistical properties of the transfer process are best captured by the distribution of the number $n$ of electron counts within a fixed time segment length $t_0$. An example of such a distribution, compiled from more than 800,000 tunneling events, is shown in Fig.\ \ref{dist}~(a). Here, $t_0=20$ ms, with a corresponding mean value $\langle\!\langle n\rangle\!\rangle$ of 17.13 electrons per time segment. The measured distribution (bars) is compared to a Poisson distribution (dashed line) with the same mean as the experimental data, a Gauss distribution (dotted line) with the same mean and variance as the experimental data, and theoretical calculations of the distribution based on a model of the experimental setup (solid line) \cite{PNAS}. The relative deviation $\delta N$ of these curves compared to the measured distribution is depicted in Fig.\ \ref{dist}~(b).

The Poisson distribution deviates significantly from the measured distribution due to its larger width. The sub-poissonian width of the measured distribution demonstrates noise suppression due to Coulomb blockade on the QD \cite{Blanter2000}. The Fano factor, i.e., the variance normalized with respect to the mean (which is 1 for a Poisson process), is reduced to nearly 0.5 due to the almost symmetric tunneling rates. For the Gauss distribution, the variance enters as an additional parameter, and it clearly fits better than the Poisson distribution. Still, however,  noticeable deviations are seen in the tails of the distribution. The best agreement with experiment, in particular in the tails of the distribution, is given by the theoretical curve derived from a two-state rate model, taking into account the finite detector bandwidth of about 40 kHz. The rates used for this calculation were determined from the waiting time distribution as described in Ref.\ \cite{PNAS}.

Next, we examine the cumulants of the distribution. As the distribution depends on the time segment length $t_0$, also the cumulants are time dependent. The cumulants of order 2 to 5, normalized with respect to the mean, are shown in Fig. \ref{cumulants}~(a). All normalized cumulants start from a value of 1 at very short times, where the transfer process is nearly poissonian.  For a Gauss distribution, only the first and second cumulants are non-zero. We observe a clear non-zero value for the third cumulant, showing that the distribution is not Gaussian, in agreement with the observed deviations in Fig. \ref{dist}. For longer times the normalized cumulants reach a constant value, referred to as the long-time limit. While the second and third cumulants show a monotonic transition between 1 and the long-time limit, the fourth and  fifth cumulants have a minimum at short times. Additionally, the fifth cumulant has a slight maximum, following the minimum, showing some indications of an oscillatory behavior. This phenomenon is further enhanced with increasing orders of the cumulants as seen in Fig.\ \ref{cumulants}~(b), showing cumulants of order $m=10$, $m=15$ and $m=20$ that clearly oscillate with time. Not only does the number of oscillations increase, but the amplitude also grows dramatically. For $m=10$, the amplitude reaches a value of 16, for $m=15$ the amplitude is around 10,000, and for  $m=20$ the amplitude exceeds 40,000,000. As a theoretical analysis has recently revealed, the magnitude of the oscillations grows factorially with the cumulant order \cite{PNAS}. More surprisingly perhaps, the oscillatory behavior of the cumulants is to be expected in a large class of nontrivial distributions as functions of almost any parameter \cite{PNAS}.

The errors on the measurement increases with the order of the cumulants and the time segment length $t_0$. This is supported by the simple error estimates in Fig.\ \ref{cumulants}~(c), where an absolute value for the estimated error is shown as functions of the cumulant order and the time segment length. The error estimates are obtained by splitting the full data set into two pieces and calculating the cumulants for both halves individually. The absolute value of the difference is then used as a simple estimate of the statistical error. The result is smoothened using a box averaging.
For increasing time an approximately exponential increase of the error is observed for each cumulant. A dotted line indicates the time at which the estimated error reaches about 10\% of the absolute value of a given cumulant. In this experiment we were able to measure the long-time limit of cumulants up to order $m=7$, which was reached after 2-3 ms.

In conclusion, we have performed measurements of the full counting statistics of electrons passing through a quantum dot. We have described our detection algorithm, shown experimental and theoretical results for the distribution functions, and presented measurements of the time evolution of the high-order cumulants together with simple error estimates. We thank T.\ Brandes, K.\ Netocn\'{y} and T.\ Novotn\'{y} for many fruitful discussions. The work was supported by the Federal Ministry of Education and Research of Germany via nanoQUIT, the German Excellence Initiative via QUEST, and the Villum Kann Rasmussen Foundation.


\begin{thebibliography}{00}

\bibitem{Elzerman2003}
J.~M. Elzerman, R.~Hanson, J.~S. Greidanus, L.~H. Willems~van Beveren,
  S.~De~Franceschi, L.~M.~K. Vandersypen, S.~Tarucha, and L.~P. Kouwenhoven,
 Phys. Rev.  B {\bf 67} (2003) 161308

\bibitem{schleser2005}
R.~Schleser, E.~Ruh, T.~Ihn, K.~Ensslin, D.~C. Driscoll, and A.~C. Gossard,
 Appl. Phys. Lett. {\bf 85} (2004) 2005

\bibitem{fujisawa2004}
T.~Fujisawa, T.~Hayashi, Y.~Hirayama, H.~D. Cheong, and Y.~H. Jeong,
 Appl. Phys. Lett. {\bf 84} (2004) 2343

\bibitem{Levitov96}
L.~S. Levitov, H.~Lee, and G.~B. Lesovik, J. Math. Phys. {\bf 37} (1996) 4845

\bibitem{Gustavsson2006}
S.~Gustavsson, R.~Leturcq, B.~Simovic, R.~Schleser, T.~Ihn, P.~Studerus,
  K.~Ensslin, D.~C. Driscoll, and A.~C. Gossard, Phys. Rev. Lett. {\bf 96} (2006) 076605

\bibitem{Fricke2007}
C.~Fricke, F.~Hohls, W.~Wegscheider, and R.~J. Haug, Phys. Rev. B  {\bf 76} (2007) 155307

\bibitem{Gustavsson2007}
S.~Gustavsson, R.~Leturcq, T.~Ihn, K.~Ensslin, M.~Reinwald, and W.~Wegscheider,
 Phys. Rev. B {\bf 75} (2007) 075314

\bibitem{PNAS}
C.~Flindt, C.~Fricke, F.~Hohls, T.\ Novotn\'{y}, K.\ Netocn\'{y}, T.~Brandes, and R.~J.
  Haug, Proc. Natl. Acad. Sci. USA {\bf 106} (2009) 10116

\bibitem{held98}
R.~Held, T.~Vancura, T.~Heinzel, K.~Ensslin, M.~Holland, and W.~Wegscheider,
 Appl. Phys. Lett.  {\bf 73} (1998) 262

\bibitem{keyser2000}
U.~F. Keyser, H.~W. Schumacher, U.~Zeitler, R.~J. Haug, and K.~Eberl,
 Appl. Phys. Lett. {\bf 76} (2000) 457

\bibitem{khrapai2006}
V.~S. Khrapai, S.~Ludwig, J.~P. Kotthaus, H.~P. Tranitz, and W.~Wegscheider,
  Phys. Rev. Lett. {\bf 97} (2006) 176803

\bibitem{Blanter2000}
Ya.~M. Blanter and M.~B\"{u}ttiker, Phys. Rep. {\bf 336} (2000) 1

\end{thebibliography}
\end{document}